\documentclass[aps, prl, twocolumn, reprint, showpacs,preprintnumbers]{revtex4-1}

\usepackage{graphicx}
\usepackage{dcolumn}
\usepackage{bm}
\usepackage{multirow}
\usepackage{color}
\usepackage{threeparttable}
\usepackage{textcomp, amsmath, amssymb}

\begin{document}

\title{Shock Response and Phase Transitions of MgO at Planetary Impact Conditions }

\author{Seth~Root}
\email{sroot@sandia.gov}
\author{Luke~Shulenburger}
\author{Raymond~W.~Lemke}
\author{Daniel~H.~Dolan}
\author{Thomas~R.~Mattsson}
\author{Michael~P.~Desjarlais}
\affiliation{Sandia National Laboratories, Albuquerque, New Mexico 87185, USA}

%
%

\date{January 30, 2015}
\preprint{{\large SAND2015-0557 O}}

\begin{abstract}
  The moon-forming impact and the subsequent evolution of the proto-Earth is strongly dependent on the properties of materials at the extreme conditions generated by this violent collision.  We examine the high pressure behavior of MgO, one of the dominant constituents in the earth's mantle, using high-precision, plate impact shock compression experiments performed on Sandia National Laboratories Z-Machine and extensive quantum simulations using Density Functional Theory (DFT) and quantum Monte Carlo (QMC).  The combined data span from ambient conditions to 1.2 TPa and 42,000 K, showing solid-solid and solid-liquid phase boundaries.  Furthermore our results indicate under impact that the solid and liquid phases coexist for more than 100 GPa, pushing complete melting to pressures in excess of 600 GPa.  The high pressure required for complete shock melting places a lower bound on the relative velocities required for the moon forming impact.
\end{abstract}


\maketitle

The leading theory of moon formation is a giant impact event occurring approximately $4.5 \times 10^{9}$ years ago \cite{Hartmann1975}.   Recent large scale simulations show that giant impacts are capable of producing the correct Earth - Moon masses and angular momentum~\cite{Canup2012,Stewart2012}.   Complicating the giant impact theory, however, is that the Earth and moon have a nearly identical chemical and isotopic composition~\cite{Wiechert2001}.  This implies that either the impactor was compositionally similar to the proto-Earth, which is thought to be unlikely, or that extensive mixing of the post impact materials occurred.  The simulations needed to test the post-impact mixing are dependent on an accurate understanding of the properties of mantle materials at extreme pressures and temperatures.  For example, post impact mixing for chemical equilibration in the proto-Lunar disk has been shown in simulations~\cite{Pahlevan2007},  but requires melting and vaporization of the mantle in order for material to diffuse.  Testing such theories is difficult because the melt line of the most common mantle materials is not well constrained at these conditions \cite{Boehler2000}. 

Advanced facilities for performing dynamic materials experiments have greatly increased the pressure and temperature regimes that can be probed for important planetary materials \cite{MDKQuartz2009, Kraus2012, Dylan2012, RootCO2}.  The Rankine-Hugoniot (RH) equations \cite{Zeldovich} relate the experimentally measured shock velocity to the thermodynamic state, pressure and density, of the shocked material.  The ability to perform experiments with steady planar shocks and with  well-characterized impactors and targets is critical for determining the equation of state (EOS) and the phase. To fully address the physics relevant to planetary science, this thermodynamic information must be augmented with an understanding of the phase transformations.  

In this work we focus on MgO, the end member of magnesiow\"ustite, a major constituent of the earth's mantle \cite{McDonough1995} and likely other terrestrial exoplanets~\cite{kepler10b}.  At ambient conditions, MgO exists in a NaCl (B1) lattice structure, which is stable over a wide pressure-temperature range~\cite{Zerr1994, DuffyDAC, Coppari2013}. Dynamic compression experiments starting from ambient~\cite{marsh1980lasl, VassiliouMgO, Duffy1994, Zhang2008, Fratanduono2013} and from T$_0$=1900~K~\cite{OlegSCCM09} show no indications of phase transitions up to 230~GPa.  \textit{Ab initio} studies on the MgO phase diagram agree that three phases exist: the B1 solid, the B2 (CsCl) solid, and the liquid~\cite{BeloMgO, boates-bonev, cebulla-MgO}, but disagree on the location of the phase boundaries. Along the Hugoniot, which is relevant for planetary impact scenarios, the location of the B1-B2 transition and the melt transition has not been precisely determined.  Recently McWilliams \textit{et al}. have shown that MgO can be dynamically compressed to pressures exceeding 1 TPa using a decaying shock technique~\cite{McWilliams2012}.  The authors also proposed locations for the B1-B2 and B2-liquid transitions along the Hugoniot, but the measurements relied heavily on prior MgO Hugoniot data, which was not well known above 200~GPa.

We present a comprehensive study of the MgO Hugoniot using experiments, density functional theory (DFT), and quantum Monte Carlo (QMC) methods over a wide pressure range covering the B1, the B2, and the liquid phases from 0.27 to 1.2~TPa.  The high-precision data constrain the Hugoniot at multi-Mbar pressures, and the DFT and QMC results further elucidate information on the phase boundaries.  This work provides accurate EOS data at extreme conditions and furthermore reveals lower limits of the relative velocity required in the giant impact scenario for moon formation.


To attain planetary impact conditions, we performed a series of shock compression experiments using the Sandia Z-Machine \cite{ZMachRef}.    The Z-machine is a pulsed power system capable of producing shaped current pulses and induced magnetic fields in excess of 20 MA and 10 MG respectively.  The combined current and magnetic field densities generate magnetic pressures up to 650 GPa that can accelerate aluminum flyers up to 40~km/s \cite{LemkeZFlyer05}.  

\begin{figure}
  \includegraphics[width=3.0in]{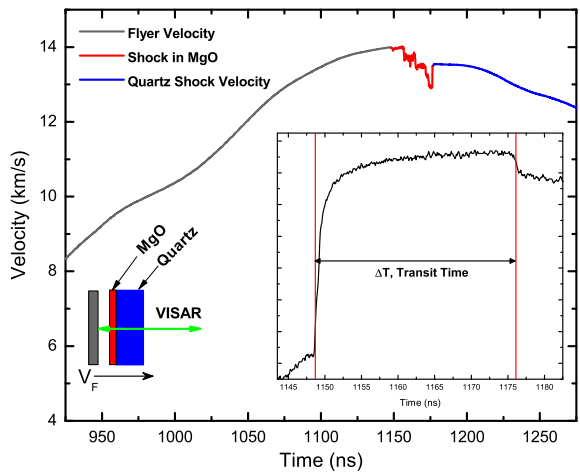}
 \caption{The experimental target and representative VISAR data. The VISAR measures the flyer velocity as it approaches the MgO (grey line).  For this low velocity impact, as the shock transits the MgO, the VISAR loses signal (red line).  As the shock transits into the quartz, the VISAR signal returns and the quartz shock velocity is measured (blue line).  The inset shows a VISAR raw signal. The fiducials correspond to impact and to shock transit into the rear window. }
 \label{ExptApproach}
 \end{figure}

Figure~\ref{ExptApproach} shows a schematic view of the target geometry; a more detailed Z target geometry is found elsewhere~\cite{RootFoam2013}.  An Al or Cu flyer plate is shocklessly accelerated toward the target stack consisting of a single-crystal MgO sample ([100], 300-500 $\mu$m, Asphera Corp., $\rho_0$ = 3.584 g/cm$^3$) backed by a window.  Although the back side of the flyer is melted by the high current, the surface of the flyer impacting the sample typically has a few hundred microns of solid density material at impact~\cite{LemkeZFlyer05}.  The velocity interferometer system for any reflector (VISAR) measures the flyer plate velocity ($V_F$) up to impact at the target (Fig.~\ref{ExptApproach}).  Impact produces a steady shock in the MgO sample.  At low impact velocities (and consequently low shock pressures), the MgO sample scatters light from the VISAR preventing direct measurement of the shock velocity.  Instead, sharp fiducials are observed in the VISAR signal (Fig.~\ref{ExptApproach} inset) that correspond to impact and to shock transit into the quartz window.   In this case, we calculated the MgO shock velocity ($U_S$) using the transit time determined from the fiducials and the measured thickness.  At high impact velocities (and high shock pressures), the shock front in the MgO is reflective and the VISAR directly measures the shock velocity.  As the shock transits into the quartz, the VISAR measures the quartz shock velocity directly.  Multiple VISAR signals were recorded for each sample eliminating 2$\pi$ ambiguities and providing redundant measurements for improved precision. For directly measured velocities, the uncertainty is better than 1\% and for transit time measurements the uncertainty is on the order of 1-2\%.

Knowing the initial densities of the MgO and the flyer plate and measuring the flyer velocity and the MgO shock velocity, we calculate the MgO Hugoniot state density ($\rho$), pressure ($P$), and particle velocity ($U_P$).  The Hugoniot state is determined using a Monte Carlo impedance matching analysis~\cite{RootCO2} to solve the RH equations~\cite{Zeldovich}.  The Monte Carlo method accounts for the uncertainties in the experimental measurement and the Al and Cu Hugoniot standards.  Further discussion on the impedance matching method along with the list of Hugoniot data are found in the supplement~\cite{supplemental-material}.

\begin{figure}	
	\includegraphics[width=3.3in]{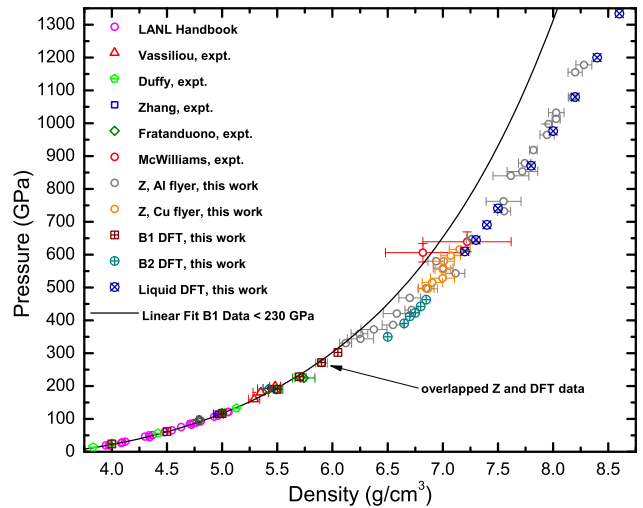}
	\caption{The MgO Hugoniot data in $\rho-P$ space from the Z experiments, previous experimental data\cite{marsh1980lasl, VassiliouMgO, Duffy1994, Zhang2008, McWilliams2012, Fratanduono2013}, and the DFT simulation results from this work. The current data set shows a deviation from the extrapolation of the fit to the B1 MgO data from $<$~230~GPa.}
	\label{rhoPdata}
\end{figure}

\begin{figure}	
	\includegraphics[width=3.3in]{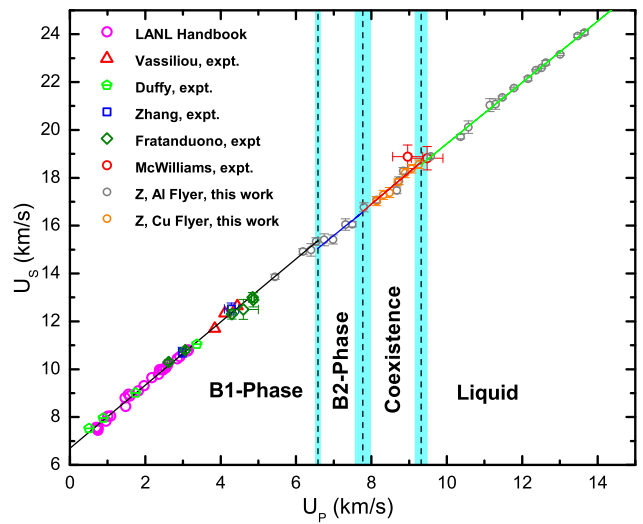}
	\caption{The experimental $U_S - U_P$ data including results from previous experiments from Refs~{\cite{marsh1980lasl, VassiliouMgO, Duffy1994, Zhang2008, McWilliams2012, Fratanduono2013}}. The optimized linear fits are also plotted.  The dashed vertical lines indicate the optimized phase boundaries along the principal Hugoniot and the shaded cyan regions indicate the uncertainty.   }
	\label{UsUpdata}
\end{figure}

Figure~\ref{rhoPdata} plots the experimental and DFT principal Hugoniot in $\rho$-P space.  The Z experimental data span the range from 0.27~TPa up to 1.2~TPa - the highest, directly measured Hugoniot states attained in MgO.  Also included are the DFT simulation results for the B1, B2, and liquid phases of MgO (discussed later).    Although the VISAR diagnostic does not give direct information about the MgO phase upon shock compression, we can infer phase transitions given our data.  Figure~\ref{rhoPdata} shows an extrapolation of the linear fit to the $U_S - U_P$ data for B1-phase Hugoniot states $<$~230~GPa (converted to $\rho$-P using the RH equations) determined from the previous experiments~\cite{marsh1980lasl, VassiliouMgO, Duffy1994, Zhang2008, Fratanduono2013}.  Below $\approx$ 360~GPa, the Z experimental data are consistent with the gas-gun data but above 360~GPa they deviate from the extrapolation.  This suggests that the B1 phase is stable up to 360~GPa and likely undergoes a phase transition from the B1 state to another phase, presumably the B2 state, at that shock pressure.  At higher pressures ($\approx > $700~GPa) the shock front in the MgO was reflective.  From the reflective shock front, we infer that the MgO has melted into a conductive fluid, similar to what is observed for quartz~\cite{MDKQuartz2009}.   

The interpretation for the MgO phase along the Hugoniot is not as obvious between the low pressure B1 phase and the high pressure liquid phase.  Prior \textit{ab initio} calculations suggest that MgO transitions from B1 to B2 and then from B2 to liquid with a small B1-B2 coexistence region and a potentially large coexistence region between the B2 and liquid phases \cite{cebulla-MgO}.  To investigate the phase regions, we analyze the $U_S - U_P$ data using a Monte Carlo optimization (MCO) method similar to the method used in previous work on carbon\cite{knudson-carbon}.   

Slope changes in the $U_S - U_P$ data often indicate phase transitions and phase boundaries. Using the MCO method, we fit four lines to the experimental $U_S - U_P$ data.  The $U_S-U_P$ data were converted to a ``cloud'' of points (as described in Ref.~\citenum{knudson-carbon}), allowing region boundaries to move smoothly during optimization.  For a particular set of data clouds, the eleven parameters (four slopes, four intercepts, and three region boundaries) of a four-segment curve were obtained by minimizing the square minimum distances to each cloud point. Revised clouds were generated by randomly drawing a new center for each cloud.  Optimization was repeated ($\approx10000$ times) using the revised clouds to characterize the distributions of the parameters.  The parameters are listed in the supplement~\cite{supplemental-material}.  It is important to note that this analysis is only possible because of the high precision data produced from the steady shocks.

Figure~\ref{UsUpdata} shows the compiled experimental $U_S - U_P$ Hugoniot data, the four linear fits, and the phase regions determined from the MCO method.  Following the literature~\cite{cebulla-MgO} and our DFT results, we propose the four regions be classified as follows: 1. The B1 solid from ambient to 363~GPa; 2. The B2 solid from 363 to 462~GPa; 3. The B2-liquid coexistence region between 462 and 620~GPa; and 4. The liquid state above 620~GPa.   However, as our continuum level experiments do not provide microstructure information, we performed \textit{ab initio} calculations of the Hugoniot and the phase diagram to better understand the high pressure states of MgO.


The high precision requirements of this work necessitated refinements of previous \textit{ab initio} methods \cite{BeloMgO,boates-bonev,cebulla-MgO}.  We performed calculations utilizing DFT and QMC focusing on the solid-solid phase transformation from B1 to B2 and the melting of MgO along the Hugoniot, presumably from the B2 phase. Using DFT to calculate the Hugoniot requires prior knowledge of the phase state, so we first calculated the phase diagram.  We used a three-part approach to determine the phase boundaries.  In order to determine the melt boundary from both the B1 and B2 phases, we performed two-phase calculations of melting using VASP 5.2.11\cite{VASPshort,PAWshort}; further details are presented in the supplemental material~\cite{supplemental-material}.  To determine the solid-solid phase boundaries we decomposed the solid's Helmholtz free energy into two pieces.
\begin{equation}
F_{sol}(V,T) = E(V) + F_{vib}(V,T)
\label{free-energy-decomp}
\end{equation}
The first piece is the density dependent energy of either the B1 or B2 phase.  This is calculated via diffusion QMC using \textsc{qmcpack}~\cite{qmcpack} following methodology detailed in Ref.~\citenum{shulenburger-prb} with particular concern paid to the construction of pseudopotentials.  The second piece of the free energy is due to the finite temperature motion of the ions and electrons and is calculated in two parts.  First the harmonic part of the free energy is calculated using the finite displacement method as implemented in the \textsc{phon} code~\cite{Alfe-phon}.   The quasiharmonic approximation (QHA) is known to break down as temperatures increase and this is particularly true for MgO~\cite{wu-anharmonic-prb}.  For this reason and because the Hugoniot is expected to cross the phase boundary relatively close to the melt line, we have augmented our QHA calculations of free energy with thermodynamic integration.  This is performed by using
\begin{equation}
\Delta S = \int_{T_i}^{T_f} \frac{1}{T} \left(\frac{\partial E}{\partial T}\right)_V dT
\end{equation}
that allows the change in entropy along an isochore to be calculated directly in terms of the internal energy.  The energy is calculated using DFT based quantum molecular dynamics (QMD) at points spaced by 250 K along several isochores in the region of the phase transition.  Using entropy from the QHA calculation at low temperatures as a reference, we calculate the Gibbs free energy of both phases and determine the phase transition pressure directly.  This method also determines the range of validity for the QHA.  We find the range to be smaller than previously estimated\cite{BeloMgO} with significant deviations in the free energy occurring by 5000 K and 400 GPa.  The positive effect of the anharmonic entropy was significantly larger in the B1 phase than in the B2 phase, moving the phase boundary to higher pressures at high temperature.  Specific computational details are in the supplement~\cite{supplemental-material}.

Figure~\ref{PhaseDiagram} shows the DFT determined phase diagram and the calculated P-T states on the MgO Hugoniot.  Using the DFT phase diagram, we can infer the microstructure during our plate impact experiments.  However, the experimental Hugoniot data is an incomplete thermodynamic description due to the lack of temperature measurements.  To assign the data to a given phase within the single phase regions, we use QMD calculations to provide the missing temperature by calculating the Hugoniot.   For each of the three phases at given densities, we perform several QMD calculations at various temperatures.  We then find the temperature for which the RH energy equation is satisfied to determine the shock state.  Finally, the pressure and temperature of these shock states are compared to the phase boundaries to determine whether that state is thermodynamically stable.  This approach also provides a validation of the QMD by comparing the calculated Hugoniot to the experimental results.   The calculations and the experiments are in good agreement in $\rho-P$ space (Fig.~\ref{rhoPdata}) and in $P-T$ space (Fig.~\ref{PhaseDiagram}).  The DFT calculated phase boundaries along the Hugoniot corroborate the MCO fitting method results for the experimental data suggesting the Hugoniot has four major regions: B1, B2, coexistence, and liquid.  Table~\ref{PhaseBounds} lists the phase boundaries along the principal Hugoniot from the MCO method and the DFT simulations.

\begin{table}
\caption{Phase boundaries along the principal Hugoniot.}
\label{PhaseBounds}
	\begin{tabular}{cccc}
	\hline
	Method		&		B1-B2 			&		B2-Coexist.				&		Coexist.-Liquid	  \\	
				&		(GPa)			&			(GPa)				&			(GPa)		\\
	\hline
	Expt. (MCO) 	&		363$\pm$6		&		462$\pm$20				&			620$\pm$17			\\
	DFT			&		330				&		475						&			600					\\
	\hline \hline
	\end{tabular}
\end{table} 


Both the experimental and DFT results show that a minimum shock pressure of 620~GPa is required to achieve complete melting in MgO that is initially at ambient temperature.  In the giant impact scenario, the proto-Earth has also been assumed to have an elevated surface temperature prior to the moon-forming event~\cite{Canup2012}.  We have performed additional DFT simulations to calculate the Hugoniot of MgO starting from an initial temperature of 1900~K.  From $T_0=1900$K, a minimum shock pressure of 445~GPa is required to achieve complete melt in the MgO.  Assuming planar normal impact, we can determine a minimum impact velocity required to melt MgO.  Table~\ref{ImpactCompare} lists the required impact velocities for impactors of common planetary materials.  In a real impact event, oblique impact \cite{Kraus2014} and shock attenuation through the mantle~\cite{Croft1982} affect the amount of energy transferred and imply even higher velocities are likely required for significant melting of MgO in the mantle.

\begin{table}
 \caption{Impactor velocities for common planetary materials required to completely melt MgO with T$_0$ at ambient and at 1900 K in a shock event (assuming planar normal impact).}
 \begin{tabular}{ccc}
   \hline
   Initial MgO Temp.   &    Impactor    &  Impact Velocity   \\
         $[K]$		&		$[300 K]$		    &      $[km/s]$  \\
   \hline
   300	&   MgO 	&  18.6    \\
   300	&   Dunite	&  19.4    \\
   300	&   Iron	&  15.3    \\
   300	&   Quartz &  20.1    \\
   \hline
   1900	&   MgO 	&    16.0	\\
   1900	&   Dunite 	&    16.3	 \\
   1900 	&   Iron   	&     12.9	\\
   1900      &   Quartz &    	17.7	\\
   \hline \hline
  \end{tabular}
   \label{ImpactCompare}
 \end{table}

We have performed an extensive experimental and computational study of the high pressure - high temperature behavior of MgO. The Hugoniot is experimentally determined up to 1.2 TPa and the data suggests that the Hugoniot crosses the B2 region with a large solid-liquid coexistence region lasting over 100 GPa before complete melt.  The DFT simulation results corroborate the findings inferred from the experimental data and also show that at elevated T$_0$ the Hugoniot persists through a large coexistence region before melt.  The results help place a lower bound on impact velocities for the moon-forming giant impact scenario.  The data and phase diagram provide a solid basis for the development of an MgO equation of state that can be used in planetary collision studies.

\begin{figure}	
	\noindent\includegraphics[width=3.1in]{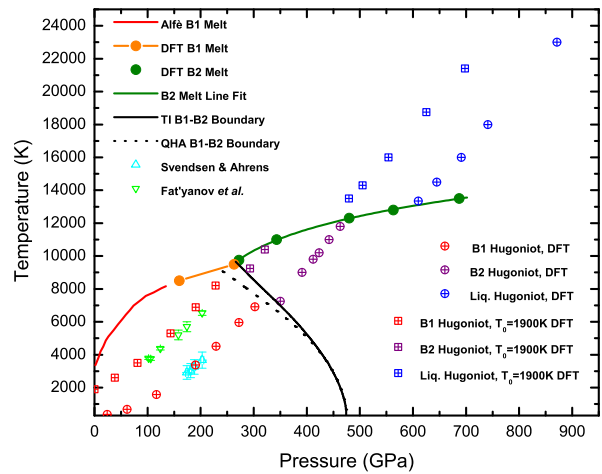}
	\caption{P-T phase diagram of MgO with calculated Hugoniots starting at ambient and elevated initial temperature conditions.  Experimental P-T data~\cite{SvendsenMgO, OlegSCCM09} and the low pressure B1-B2 melt line from Ref.~\citenum{Alfe-MgO-2phase} are included.}
	\label{PhaseDiagram}
\end{figure}

The authors thank the Z-Operations and Fabrication team for assembling targets and fielding the Z experiments.  The authors also thank K. Cochrane for insightful discussions.  QMC calculations by LS were supported through the Predictive Theory and Modeling for Materials and Chemical Science program by the Office of Basic Energy Science (BES), Department of Energy (DOE). Sandia National Laboratories is a multiprogram laboratory managed and operated by Sandia Corporation, a wholly owned subsidiary of Lockheed Martin Corporation, for the U.S. Department of Energy's National Nuclear Security Administration under Contract No. DE-AC04-94AL85000.

\bibliographystyle{apsrev4-1}

\begin{thebibliography}{41}%
\makeatletter
\providecommand \@ifxundefined [1]{%
 \@ifx{#1\undefined}
}%
\providecommand \@ifnum [1]{%
 \ifnum #1\expandafter \@firstoftwo
 \else \expandafter \@secondoftwo
 \fi
}%
\providecommand \@ifx [1]{%
 \ifx #1\expandafter \@firstoftwo
 \else \expandafter \@secondoftwo
 \fi
}%
\providecommand \natexlab [1]{#1}%
\providecommand \enquote  [1]{``#1''}%
\providecommand \bibnamefont  [1]{#1}%
\providecommand \bibfnamefont [1]{#1}%
\providecommand \citenamefont [1]{#1}%
\providecommand \href@noop [0]{\@secondoftwo}%
\providecommand \href [0]{\begingroup \@sanitize@url \@href}%
\providecommand \@href[1]{\@@startlink{#1}\@@href}%
\providecommand \@@href[1]{\endgroup#1\@@endlink}%
\providecommand \@sanitize@url [0]{\catcode `\\12\catcode `\$12\catcode
  `\&12\catcode `\#12\catcode `\^12\catcode `\_12\catcode `\%12\relax}%
\providecommand \@@startlink[1]{}%
\providecommand \@@endlink[0]{}%
\providecommand \url  [0]{\begingroup\@sanitize@url \@url }%
\providecommand \@url [1]{\endgroup\@href {#1}{\urlprefix }}%
\providecommand \urlprefix  [0]{URL }%
\providecommand \Eprint [0]{\href }%
\providecommand \doibase [0]{http://dx.doi.org/}%
\providecommand \selectlanguage [0]{\@gobble}%
\providecommand \bibinfo  [0]{\@secondoftwo}%
\providecommand \bibfield  [0]{\@secondoftwo}%
\providecommand \translation [1]{[#1]}%
\providecommand \BibitemOpen [0]{}%
\providecommand \bibitemStop [0]{}%
\providecommand \bibitemNoStop [0]{.\EOS\space}%
\providecommand \EOS [0]{\spacefactor3000\relax}%
\providecommand \BibitemShut  [1]{\csname bibitem#1\endcsname}%
\let\auto@bib@innerbib\@empty
\bibitem [{\citenamefont {Hartmann}\ and\ \citenamefont
  {Davis}(1975)}]{Hartmann1975}%
  \BibitemOpen
  \bibfield  {author} {\bibinfo {author} {\bibfnamefont {W.~K.}\ \bibnamefont
  {Hartmann}}\ and\ \bibinfo {author} {\bibfnamefont {D.~R.}\ \bibnamefont
  {Davis}},\ }\href@noop {} {\bibfield  {journal} {\bibinfo  {journal}
  {Icarus}\ }\textbf {\bibinfo {volume} {24}},\ \bibinfo {pages} {504}
  (\bibinfo {year} {1975})}\BibitemShut {NoStop}%
\bibitem [{\citenamefont {Canup}(2012)}]{Canup2012}%
  \BibitemOpen
  \bibfield  {author} {\bibinfo {author} {\bibfnamefont {R.~M.}\ \bibnamefont
  {Canup}},\ }\href@noop {} {\bibfield  {journal} {\bibinfo  {journal}
  {Science}\ }\textbf {\bibinfo {volume} {338}},\ \bibinfo {pages} {1052}
  (\bibinfo {year} {2012})}\BibitemShut {NoStop}%
\bibitem [{\citenamefont {Cuk}\ and\ \citenamefont
  {Stewart}(2012)}]{Stewart2012}%
  \BibitemOpen
  \bibfield  {author} {\bibinfo {author} {\bibfnamefont {M.}~\bibnamefont
  {Cuk}}\ and\ \bibinfo {author} {\bibfnamefont {S.~T.}\ \bibnamefont
  {Stewart}},\ }\href@noop {} {\bibfield  {journal} {\bibinfo  {journal}
  {Science}\ }\textbf {\bibinfo {volume} {338}},\ \bibinfo {pages} {1047}
  (\bibinfo {year} {2012})}\BibitemShut {NoStop}%
\bibitem [{\citenamefont {Wiechert}\ \emph {et~al.}(2001)\citenamefont
  {Wiechert}, \citenamefont {Halliday}, \citenamefont {Lee}, \citenamefont
  {Snyder}, \citenamefont {Taylor},\ and\ \citenamefont
  {Rumble}}]{Wiechert2001}%
  \BibitemOpen
  \bibfield  {author} {\bibinfo {author} {\bibfnamefont {U.}~\bibnamefont
  {Wiechert}}, \bibinfo {author} {\bibfnamefont {A.~N.}\ \bibnamefont
  {Halliday}}, \bibinfo {author} {\bibfnamefont {D.-C.}\ \bibnamefont {Lee}},
  \bibinfo {author} {\bibfnamefont {G.~A.}\ \bibnamefont {Snyder}}, \bibinfo
  {author} {\bibfnamefont {L.~A.}\ \bibnamefont {Taylor}}, \ and\ \bibinfo
  {author} {\bibfnamefont {D.}~\bibnamefont {Rumble}},\ }\href@noop {}
  {\bibfield  {journal} {\bibinfo  {journal} {Science}\ }\textbf {\bibinfo
  {volume} {294}},\ \bibinfo {pages} {345} (\bibinfo {year}
  {2001})}\BibitemShut {NoStop}%
\bibitem [{\citenamefont {Pahlevan}\ and\ \citenamefont
  {Stevenson}(2007)}]{Pahlevan2007}%
  \BibitemOpen
  \bibfield  {author} {\bibinfo {author} {\bibfnamefont {K.}~\bibnamefont
  {Pahlevan}}\ and\ \bibinfo {author} {\bibfnamefont {D.~J.}\ \bibnamefont
  {Stevenson}},\ }\href@noop {} {\bibfield  {journal} {\bibinfo  {journal}
  {Earth Planet. Sci. Lett.}\ }\textbf {\bibinfo {volume} {262}},\ \bibinfo
  {pages} {438} (\bibinfo {year} {2007})}\BibitemShut {NoStop}%
\bibitem [{\citenamefont {Boehler}(2000)}]{Boehler2000}%
  \BibitemOpen
  \bibfield  {author} {\bibinfo {author} {\bibfnamefont {R.}~\bibnamefont
  {Boehler}},\ }\href@noop {} {\bibfield  {journal} {\bibinfo  {journal} {Rev.
  Geophysics}\ }\textbf {\bibinfo {volume} {38}},\ \bibinfo {pages} {221}
  (\bibinfo {year} {2000})}\BibitemShut {NoStop}%
\bibitem [{\citenamefont {Knudson}\ and\ \citenamefont
  {Desjarlais}(2009)}]{MDKQuartz2009}%
  \BibitemOpen
  \bibfield  {author} {\bibinfo {author} {\bibfnamefont {M.~D.}\ \bibnamefont
  {Knudson}}\ and\ \bibinfo {author} {\bibfnamefont {M.~P.}\ \bibnamefont
  {Desjarlais}},\ }\href@noop {} {\bibfield  {journal} {\bibinfo  {journal}
  {Phys. Rev. Lett.}\ }\textbf {\bibinfo {volume} {103}},\ \bibinfo {pages}
  {225501} (\bibinfo {year} {2009})}\BibitemShut {NoStop}%
\bibitem [{\citenamefont {Kraus}\ \emph {et~al.}(2012)\citenamefont {Kraus},
  \citenamefont {Stewart}, \citenamefont {Swift}, \citenamefont {Bolme},
  \citenamefont {Smith}, \citenamefont {Hamel}, \citenamefont {Hammel},
  \citenamefont {Spaulding}, \citenamefont {Hicks}, \citenamefont {Eggert},\
  and\ \citenamefont {Collins}}]{Kraus2012}%
  \BibitemOpen
  \bibfield  {author} {\bibinfo {author} {\bibfnamefont {R.~G.}\ \bibnamefont
  {Kraus}}, \bibinfo {author} {\bibfnamefont {S.~T.}\ \bibnamefont {Stewart}},
  \bibinfo {author} {\bibfnamefont {D.~C.}\ \bibnamefont {Swift}}, \bibinfo
  {author} {\bibfnamefont {C.~A.}\ \bibnamefont {Bolme}}, \bibinfo {author}
  {\bibfnamefont {R.~F.}\ \bibnamefont {Smith}}, \bibinfo {author}
  {\bibfnamefont {S.}~\bibnamefont {Hamel}}, \bibinfo {author} {\bibfnamefont
  {B.~D.}\ \bibnamefont {Hammel}}, \bibinfo {author} {\bibfnamefont {D.~K.}\
  \bibnamefont {Spaulding}}, \bibinfo {author} {\bibfnamefont {D.~G.}\
  \bibnamefont {Hicks}}, \bibinfo {author} {\bibfnamefont {J.~H.}\ \bibnamefont
  {Eggert}}, \ and\ \bibinfo {author} {\bibfnamefont {G.~W.}\ \bibnamefont
  {Collins}},\ }\href@noop {} {\bibfield  {journal} {\bibinfo  {journal} {J.
  Geophys. Res. - Planets}\ }\textbf {\bibinfo {volume} {117}},\ \bibinfo
  {pages} {E09009} (\bibinfo {year} {2012})}\BibitemShut {NoStop}%
\bibitem [{\citenamefont {Spaulding}\ \emph {et~al.}(2012)\citenamefont
  {Spaulding}, \citenamefont {McWilliams}, \citenamefont {Jeanloz},
  \citenamefont {Eggert}, \citenamefont {Celliers}, \citenamefont {Hicks},
  \citenamefont {Collins},\ and\ \citenamefont {Smith}}]{Dylan2012}%
  \BibitemOpen
  \bibfield  {author} {\bibinfo {author} {\bibfnamefont {D.~K.}\ \bibnamefont
  {Spaulding}}, \bibinfo {author} {\bibfnamefont {R.~S.}\ \bibnamefont
  {McWilliams}}, \bibinfo {author} {\bibfnamefont {R.}~\bibnamefont {Jeanloz}},
  \bibinfo {author} {\bibfnamefont {J.~H.}\ \bibnamefont {Eggert}}, \bibinfo
  {author} {\bibfnamefont {P.~M.}\ \bibnamefont {Celliers}}, \bibinfo {author}
  {\bibfnamefont {D.~G.}\ \bibnamefont {Hicks}}, \bibinfo {author}
  {\bibfnamefont {G.~W.}\ \bibnamefont {Collins}}, \ and\ \bibinfo {author}
  {\bibfnamefont {R.~F.}\ \bibnamefont {Smith}},\ }\href@noop {} {\bibfield
  {journal} {\bibinfo  {journal} {Phys. Rev. Lett.}\ }\textbf {\bibinfo
  {volume} {108}},\ \bibinfo {pages} {065701} (\bibinfo {year}
  {2012})}\BibitemShut {NoStop}%
\bibitem [{\citenamefont {Root}\ \emph
  {et~al.}(2013{\natexlab{a}})\citenamefont {Root}, \citenamefont {Cochrane},
  \citenamefont {Carpenter},\ and\ \citenamefont {Mattsson}}]{RootCO2}%
  \BibitemOpen
  \bibfield  {author} {\bibinfo {author} {\bibfnamefont {S.}~\bibnamefont
  {Root}}, \bibinfo {author} {\bibfnamefont {K.~R.}\ \bibnamefont {Cochrane}},
  \bibinfo {author} {\bibfnamefont {J.~H.}\ \bibnamefont {Carpenter}}, \ and\
  \bibinfo {author} {\bibfnamefont {T.~R.}\ \bibnamefont {Mattsson}},\
  }\href@noop {} {\bibfield  {journal} {\bibinfo  {journal} {Phys. Rev. B}\
  }\textbf {\bibinfo {volume} {87}},\ \bibinfo {pages} {224102} (\bibinfo
  {year} {2013}{\natexlab{a}})}\BibitemShut {NoStop}%
\bibitem [{\citenamefont {Zel'Dovich}\ and\ \citenamefont
  {Raizer}(2002)}]{Zeldovich}%
  \BibitemOpen
  \bibfield  {author} {\bibinfo {author} {\bibfnamefont {Y.~B.}\ \bibnamefont
  {Zel'Dovich}}\ and\ \bibinfo {author} {\bibfnamefont {Y.~P.}\ \bibnamefont
  {Raizer}},\ }\href@noop {} {\emph {\bibinfo {title} {Physics of Shock Waves
  and High Temperature Phenomena}}}\ (\bibinfo  {publisher} {Dover
  Publications, Inc.},\ \bibinfo {address} {Mineola, NY},\ \bibinfo {year}
  {2002})\BibitemShut {NoStop}%
\bibitem [{\citenamefont {McDonough}\ and\ \citenamefont
  {Sun}(1995)}]{McDonough1995}%
  \BibitemOpen
  \bibfield  {author} {\bibinfo {author} {\bibfnamefont {W.~F.}\ \bibnamefont
  {McDonough}}\ and\ \bibinfo {author} {\bibfnamefont {S.}~\bibnamefont
  {Sun}},\ }\href@noop {} {\bibfield  {journal} {\bibinfo  {journal} {Chem.
  Geology}\ }\textbf {\bibinfo {volume} {120}},\ \bibinfo {pages} {223}
  (\bibinfo {year} {1995})}\BibitemShut {NoStop}%
\bibitem [{\citenamefont {Batalha}\ \emph {et~al.}(2011)\citenamefont {Batalha}
  \emph {et~al.}}]{kepler10b}%
  \BibitemOpen
  \bibfield  {author} {\bibinfo {author} {\bibfnamefont {N.~M.}\ \bibnamefont
  {Batalha}} \emph {et~al.},\ }\href@noop {} {\bibfield  {journal} {\bibinfo
  {journal} {Astrophys. J.}\ }\textbf {\bibinfo {volume} {27}},\ \bibinfo
  {pages} {729} (\bibinfo {year} {2011})}\BibitemShut {NoStop}%
\bibitem [{\citenamefont {Zerr}\ and\ \citenamefont
  {Boehler}(1994)}]{Zerr1994}%
  \BibitemOpen
  \bibfield  {author} {\bibinfo {author} {\bibfnamefont {A.}~\bibnamefont
  {Zerr}}\ and\ \bibinfo {author} {\bibfnamefont {R.}~\bibnamefont {Boehler}},\
  }\href@noop {} {\bibfield  {journal} {\bibinfo  {journal} {Nature}\ }\textbf
  {\bibinfo {volume} {371}},\ \bibinfo {pages} {506} (\bibinfo {year}
  {1994})}\BibitemShut {NoStop}%
\bibitem [{\citenamefont {Duffy}\ \emph {et~al.}(1995)\citenamefont {Duffy},
  \citenamefont {Hemley},\ and\ \citenamefont {Mao}}]{DuffyDAC}%
  \BibitemOpen
  \bibfield  {author} {\bibinfo {author} {\bibfnamefont {T.~S.}\ \bibnamefont
  {Duffy}}, \bibinfo {author} {\bibfnamefont {R.~J.}\ \bibnamefont {Hemley}}, \
  and\ \bibinfo {author} {\bibfnamefont {H.}~\bibnamefont {Mao}},\ }\href@noop
  {} {\bibfield  {journal} {\bibinfo  {journal} {Phys. Rev. Lett.}\ }\textbf
  {\bibinfo {volume} {74}},\ \bibinfo {pages} {1371} (\bibinfo {year}
  {1995})}\BibitemShut {NoStop}%
\bibitem [{\citenamefont {Coppari}\ \emph {et~al.}(2013)\citenamefont
  {Coppari}, \citenamefont {Smith}, \citenamefont {Eggert}, \citenamefont
  {Wang}, \citenamefont {Rygg}, \citenamefont {Lazicki}, \citenamefont
  {Hawreliak}, \citenamefont {Collins},\ and\ \citenamefont
  {Duffy}}]{Coppari2013}%
  \BibitemOpen
  \bibfield  {author} {\bibinfo {author} {\bibfnamefont {F.}~\bibnamefont
  {Coppari}}, \bibinfo {author} {\bibfnamefont {R.~F.}\ \bibnamefont {Smith}},
  \bibinfo {author} {\bibfnamefont {J.~H.}\ \bibnamefont {Eggert}}, \bibinfo
  {author} {\bibfnamefont {J.}~\bibnamefont {Wang}}, \bibinfo {author}
  {\bibfnamefont {J.~R.}\ \bibnamefont {Rygg}}, \bibinfo {author}
  {\bibfnamefont {A.}~\bibnamefont {Lazicki}}, \bibinfo {author} {\bibfnamefont
  {J.~A.}\ \bibnamefont {Hawreliak}}, \bibinfo {author} {\bibfnamefont {G.~W.}\
  \bibnamefont {Collins}}, \ and\ \bibinfo {author} {\bibfnamefont {T.~S.}\
  \bibnamefont {Duffy}},\ }\href@noop {} {\bibfield  {journal} {\bibinfo
  {journal} {Nature Geo.}\ }\textbf {\bibinfo {volume} {6}},\ \bibinfo {pages}
  {926} (\bibinfo {year} {2013})}\BibitemShut {NoStop}%
\bibitem [{\citenamefont {Marsh}(1980)}]{marsh1980lasl}%
  \BibitemOpen
  \bibfield  {author} {\bibinfo {author} {\bibfnamefont {S.~P.}\ \bibnamefont
  {Marsh}},\ }\href@noop {} {\emph {\bibinfo {title} {LASL Shock Hugoniot
  Data}}},\ Vol.~\bibinfo {volume} {5}\ (\bibinfo  {publisher} {Univ of
  California Press},\ \bibinfo {year} {1980})\BibitemShut {NoStop}%
\bibitem [{\citenamefont {Vassiliou}\ and\ \citenamefont
  {Ahrens}(1981)}]{VassiliouMgO}%
  \BibitemOpen
  \bibfield  {author} {\bibinfo {author} {\bibfnamefont {M.~S.}\ \bibnamefont
  {Vassiliou}}\ and\ \bibinfo {author} {\bibfnamefont {T.~J.}\ \bibnamefont
  {Ahrens}},\ }\href@noop {} {\bibfield  {journal} {\bibinfo  {journal}
  {Geophys. Res. Lett.}\ }\textbf {\bibinfo {volume} {8}},\ \bibinfo {pages}
  {729} (\bibinfo {year} {1981})}\BibitemShut {NoStop}%
\bibitem [{\citenamefont {Duffy}\ and\ \citenamefont
  {Ahrens}(1994)}]{Duffy1994}%
  \BibitemOpen
  \bibfield  {author} {\bibinfo {author} {\bibfnamefont {T.~S.}\ \bibnamefont
  {Duffy}}\ and\ \bibinfo {author} {\bibfnamefont {T.~J.}\ \bibnamefont
  {Ahrens}},\ }\href@noop {} {\bibfield  {journal} {\bibinfo  {journal} {AIP
  Conference Proceedings}\ }\textbf {\bibinfo {volume} {309}},\ \bibinfo
  {pages} {1107} (\bibinfo {year} {1994})}\BibitemShut {NoStop}%
\bibitem [{\citenamefont {Zhang}\ \emph {et~al.}(2008)\citenamefont {Zhang},
  \citenamefont {Gong},\ and\ \citenamefont {Fei}}]{Zhang2008}%
  \BibitemOpen
  \bibfield  {author} {\bibinfo {author} {\bibfnamefont {L.}~\bibnamefont
  {Zhang}}, \bibinfo {author} {\bibfnamefont {Z.}~\bibnamefont {Gong}}, \ and\
  \bibinfo {author} {\bibfnamefont {Y.}~\bibnamefont {Fei}},\ }\href@noop {}
  {\bibfield  {journal} {\bibinfo  {journal} {J. Phys. Chem. Solids}\ }\textbf
  {\bibinfo {volume} {69}},\ \bibinfo {pages} {2344} (\bibinfo {year}
  {2008})}\BibitemShut {NoStop}%
\bibitem [{\citenamefont {Fratanduono}\ \emph {et~al.}(2013)\citenamefont
  {Fratanduono}, \citenamefont {Eggert}, \citenamefont {Akin}, \citenamefont
  {Chau},\ and\ \citenamefont {Holmes}}]{Fratanduono2013}%
  \BibitemOpen
  \bibfield  {author} {\bibinfo {author} {\bibfnamefont {D.~E.}\ \bibnamefont
  {Fratanduono}}, \bibinfo {author} {\bibfnamefont {J.~H.}\ \bibnamefont
  {Eggert}}, \bibinfo {author} {\bibfnamefont {M.~C.}\ \bibnamefont {Akin}},
  \bibinfo {author} {\bibfnamefont {R.}~\bibnamefont {Chau}}, \ and\ \bibinfo
  {author} {\bibfnamefont {N.~C.}\ \bibnamefont {Holmes}},\ }\href@noop {}
  {\bibfield  {journal} {\bibinfo  {journal} {J. Appl. Phys.}\ }\textbf
  {\bibinfo {volume} {114}},\ \bibinfo {pages} {043518} (\bibinfo {year}
  {2013})}\BibitemShut {NoStop}%
\bibitem [{\citenamefont {Fat'yanov}\ \emph {et~al.}(2009)\citenamefont
  {Fat'yanov}, \citenamefont {Asimow},\ and\ \citenamefont
  {Ahrens}}]{OlegSCCM09}%
  \BibitemOpen
  \bibfield  {author} {\bibinfo {author} {\bibfnamefont {O.~V.}\ \bibnamefont
  {Fat'yanov}}, \bibinfo {author} {\bibfnamefont {P.~D.}\ \bibnamefont
  {Asimow}}, \ and\ \bibinfo {author} {\bibfnamefont {T.~J.}\ \bibnamefont
  {Ahrens}},\ }in\ \href@noop {} {\emph {\bibinfo {booktitle} {Shock
  Compression of Condensed Matter}}},\ \bibinfo {editor} {edited by\ \bibinfo
  {editor} {\bibfnamefont {M.~L.}\ \bibnamefont {Elert}}, \bibinfo {editor}
  {\bibfnamefont {W.~T. B. M.~D.}\ \bibnamefont {Furnish}}, \bibinfo {editor}
  {\bibfnamefont {W.~W.}\ \bibnamefont {Anderson}}, \ and\ \bibinfo {editor}
  {\bibfnamefont {W.~G.}\ \bibnamefont {Proud}}}\ (\bibinfo  {publisher}
  {AIP},\ \bibinfo {year} {2009})\ p.\ \bibinfo {pages} {855}\BibitemShut
  {NoStop}%
\bibitem [{\citenamefont {Belonoshko}\ \emph {et~al.}(2010)\citenamefont
  {Belonoshko}, \citenamefont {Arapan}, \citenamefont {Martonak},\ and\
  \citenamefont {Rosengren}}]{BeloMgO}%
  \BibitemOpen
  \bibfield  {author} {\bibinfo {author} {\bibfnamefont {A.~B.}\ \bibnamefont
  {Belonoshko}}, \bibinfo {author} {\bibfnamefont {S.}~\bibnamefont {Arapan}},
  \bibinfo {author} {\bibfnamefont {R.}~\bibnamefont {Martonak}}, \ and\
  \bibinfo {author} {\bibfnamefont {A.}~\bibnamefont {Rosengren}},\ }\href@noop
  {} {\bibfield  {journal} {\bibinfo  {journal} {Phys. Rev. B}\ }\textbf
  {\bibinfo {volume} {81}},\ \bibinfo {pages} {054110} (\bibinfo {year}
  {2010})}\BibitemShut {NoStop}%
\bibitem [{\citenamefont {Boates}\ and\ \citenamefont
  {Bonev}(2013)}]{boates-bonev}%
  \BibitemOpen
  \bibfield  {author} {\bibinfo {author} {\bibfnamefont {B.}~\bibnamefont
  {Boates}}\ and\ \bibinfo {author} {\bibfnamefont {S.~A.}\ \bibnamefont
  {Bonev}},\ }\href@noop {} {\bibfield  {journal} {\bibinfo  {journal} {Phys.
  Rev. Lett.}\ }\textbf {\bibinfo {volume} {110}},\ \bibinfo {pages} {135504}
  (\bibinfo {year} {2013})}\BibitemShut {NoStop}%
\bibitem [{\citenamefont {Cebulla}\ and\ \citenamefont
  {Redmer}(2014)}]{cebulla-MgO}%
  \BibitemOpen
  \bibfield  {author} {\bibinfo {author} {\bibfnamefont {D.}~\bibnamefont
  {Cebulla}}\ and\ \bibinfo {author} {\bibfnamefont {R.}~\bibnamefont
  {Redmer}},\ }\href@noop {} {\bibfield  {journal} {\bibinfo  {journal} {Phys.
  Rev. B}\ }\textbf {\bibinfo {volume} {89}},\ \bibinfo {pages} {134107}
  (\bibinfo {year} {2014})}\BibitemShut {NoStop}%
\bibitem [{\citenamefont {McWilliams}\ \emph {et~al.}(2012)\citenamefont
  {McWilliams}, \citenamefont {Spaulding}, \citenamefont {Eggert},
  \citenamefont {Celliers}, \citenamefont {Hicks}, \citenamefont {Smith},
  \citenamefont {Collins},\ and\ \citenamefont {Jeanloz}}]{McWilliams2012}%
  \BibitemOpen
  \bibfield  {author} {\bibinfo {author} {\bibfnamefont {R.~S.}\ \bibnamefont
  {McWilliams}}, \bibinfo {author} {\bibfnamefont {D.~K.}\ \bibnamefont
  {Spaulding}}, \bibinfo {author} {\bibfnamefont {J.~H.}\ \bibnamefont
  {Eggert}}, \bibinfo {author} {\bibfnamefont {P.~M.}\ \bibnamefont
  {Celliers}}, \bibinfo {author} {\bibfnamefont {D.~G.}\ \bibnamefont {Hicks}},
  \bibinfo {author} {\bibfnamefont {R.~F.}\ \bibnamefont {Smith}}, \bibinfo
  {author} {\bibfnamefont {G.~W.}\ \bibnamefont {Collins}}, \ and\ \bibinfo
  {author} {\bibfnamefont {R.}~\bibnamefont {Jeanloz}},\ }\href@noop {}
  {\bibfield  {journal} {\bibinfo  {journal} {Science}\ }\textbf {\bibinfo
  {volume} {338}},\ \bibinfo {pages} {1330} (\bibinfo {year}
  {2012})}\BibitemShut {NoStop}%
\bibitem [{\citenamefont {{M. E. Savage \textit{et al.}}}(2007)}]{ZMachRef}%
  \BibitemOpen
  \bibfield  {author} {\bibinfo {author} {\bibnamefont {{M. E. Savage
  \textit{et al.}}}},\ }in\ \href@noop {} {\emph {\bibinfo {booktitle} {2007
  IEEE Pulsed Power Conference}}},\ Vol.\ \bibinfo {volume} {1-4}\ (\bibinfo
  {year} {2007})\ p.\ \bibinfo {pages} {979}\BibitemShut {NoStop}%
\bibitem [{\citenamefont {Lemke}\ \emph {et~al.}(2005)\citenamefont {Lemke},
  \citenamefont {Knudson}, \citenamefont {Bliss}, \citenamefont {Cochrane},
  \citenamefont {Davis}, \citenamefont {Giunta}, \citenamefont {Harjes},\ and\
  \citenamefont {Slutz}}]{LemkeZFlyer05}%
  \BibitemOpen
  \bibfield  {author} {\bibinfo {author} {\bibfnamefont {R.~W.}\ \bibnamefont
  {Lemke}}, \bibinfo {author} {\bibfnamefont {M.~D.}\ \bibnamefont {Knudson}},
  \bibinfo {author} {\bibfnamefont {D.~E.}\ \bibnamefont {Bliss}}, \bibinfo
  {author} {\bibfnamefont {K.}~\bibnamefont {Cochrane}}, \bibinfo {author}
  {\bibfnamefont {J.-P.}\ \bibnamefont {Davis}}, \bibinfo {author}
  {\bibfnamefont {A.~A.}\ \bibnamefont {Giunta}}, \bibinfo {author}
  {\bibfnamefont {H.~C.}\ \bibnamefont {Harjes}}, \ and\ \bibinfo {author}
  {\bibfnamefont {S.~A.}\ \bibnamefont {Slutz}},\ }\href@noop {} {\bibfield
  {journal} {\bibinfo  {journal} {J. Appl. Phys.}\ }\textbf {\bibinfo {volume}
  {98}},\ \bibinfo {pages} {073530} (\bibinfo {year} {2005})}\BibitemShut
  {NoStop}%
\bibitem [{\citenamefont {Root}\ \emph
  {et~al.}(2013{\natexlab{b}})\citenamefont {Root}, \citenamefont {Haill},
  \citenamefont {Lane}, \citenamefont {Thompson}, \citenamefont {Grest},
  \citenamefont {Schroen},\ and\ \citenamefont {Mattsson}}]{RootFoam2013}%
  \BibitemOpen
  \bibfield  {author} {\bibinfo {author} {\bibfnamefont {S.}~\bibnamefont
  {Root}}, \bibinfo {author} {\bibfnamefont {T.~A.}\ \bibnamefont {Haill}},
  \bibinfo {author} {\bibfnamefont {J.~M.~D.}\ \bibnamefont {Lane}}, \bibinfo
  {author} {\bibfnamefont {A.~P.}\ \bibnamefont {Thompson}}, \bibinfo {author}
  {\bibfnamefont {G.~S.}\ \bibnamefont {Grest}}, \bibinfo {author}
  {\bibfnamefont {D.~G.}\ \bibnamefont {Schroen}}, \ and\ \bibinfo {author}
  {\bibfnamefont {T.~R.}\ \bibnamefont {Mattsson}},\ }\href@noop {} {\bibfield
  {journal} {\bibinfo  {journal} {J. Appl. Phys.}\ }\textbf {\bibinfo {volume}
  {114}},\ \bibinfo {pages} {103502} (\bibinfo {year}
  {2013}{\natexlab{b}})}\BibitemShut {NoStop}%
\bibitem [{sup()}]{supplemental-material}%
  \BibitemOpen
  \href@noop {} {}\bibinfo {note} {See Supplemental Material at [URL will be
  inserted by publisher]}\BibitemShut {NoStop}%
\bibitem [{\citenamefont {Knudson}\ \emph {et~al.}(2008)\citenamefont
  {Knudson}, \citenamefont {Desjarlais},\ and\ \citenamefont
  {Dolan}}]{knudson-carbon}%
  \BibitemOpen
  \bibfield  {author} {\bibinfo {author} {\bibfnamefont {M.~D.}\ \bibnamefont
  {Knudson}}, \bibinfo {author} {\bibfnamefont {M.~P.}\ \bibnamefont
  {Desjarlais}}, \ and\ \bibinfo {author} {\bibfnamefont {D.~H.}\ \bibnamefont
  {Dolan}},\ }\href@noop {} {\bibfield  {journal} {\bibinfo  {journal}
  {Science}\ }\textbf {\bibinfo {volume} {322}},\ \bibinfo {pages} {1822}
  (\bibinfo {year} {2008})}\BibitemShut {NoStop}%
\bibitem [{VAS()}]{VASPshort}%
  \BibitemOpen
  \href@noop {} {}\bibinfo {note} {G. Kresse and J. Hafner, Phys. Rev. B {\bf
  47}, R558 (1993), Phys.~Rev.~B {\bf 49}, 14251 (1994); G. Kresse and J.
  Furthm\"{u}ller, Phys. Rev. B {\bf 54}, 11169 (1996).}\BibitemShut {Stop}%
\bibitem [{PAW()}]{PAWshort}%
  \BibitemOpen
  \href@noop {} {}\bibinfo {note} {P. E. Bl\"ochl, Phys. Rev. B {\bf 50}, 17953
  (1994); G. Kresse and D. Joubert, Phys. Rev. B {\bf 59}, 1758
  (1999).}\BibitemShut {Stop}%
\bibitem [{\citenamefont {Kim}\ \emph {et~al.}(2012)\citenamefont {Kim},
  \citenamefont {Esler}, \citenamefont {McMinis}, \citenamefont {Morales},
  \citenamefont {Clark}, \citenamefont {Shulenburger},\ and\ \citenamefont
  {Ceperley}}]{qmcpack}%
  \BibitemOpen
  \bibfield  {author} {\bibinfo {author} {\bibfnamefont {J.}~\bibnamefont
  {Kim}}, \bibinfo {author} {\bibfnamefont {K.~P.}\ \bibnamefont {Esler}},
  \bibinfo {author} {\bibfnamefont {J.}~\bibnamefont {McMinis}}, \bibinfo
  {author} {\bibfnamefont {M.~A.}\ \bibnamefont {Morales}}, \bibinfo {author}
  {\bibfnamefont {B.~K.}\ \bibnamefont {Clark}}, \bibinfo {author}
  {\bibfnamefont {L.}~\bibnamefont {Shulenburger}}, \ and\ \bibinfo {author}
  {\bibfnamefont {D.~M.}\ \bibnamefont {Ceperley}},\ }\href@noop {} {\bibfield
  {journal} {\bibinfo  {journal} {J. Phys. Conf. Series}\ }\textbf {\bibinfo
  {volume} {402}},\ \bibinfo {pages} {012008} (\bibinfo {year}
  {2012})}\BibitemShut {NoStop}%
\bibitem [{\citenamefont {Shulenburger}\ and\ \citenamefont
  {Mattsson}(2013)}]{shulenburger-prb}%
  \BibitemOpen
  \bibfield  {author} {\bibinfo {author} {\bibfnamefont {L.}~\bibnamefont
  {Shulenburger}}\ and\ \bibinfo {author} {\bibfnamefont {T.~R.}\ \bibnamefont
  {Mattsson}},\ }\href@noop {} {\bibfield  {journal} {\bibinfo  {journal}
  {Phys. Rev. B}\ }\textbf {\bibinfo {volume} {88}},\ \bibinfo {pages} {245117}
  (\bibinfo {year} {2013})}\BibitemShut {NoStop}%
\bibitem [{\citenamefont {Alf{\`e}}(2009)}]{Alfe-phon}%
  \BibitemOpen
  \bibfield  {author} {\bibinfo {author} {\bibfnamefont {D.}~\bibnamefont
  {Alf{\`e}}},\ }\href@noop {} {\bibfield  {journal} {\bibinfo  {journal}
  {Computer Physics Communications}\ }\textbf {\bibinfo {volume} {180}},\
  \bibinfo {pages} {2622} (\bibinfo {year} {2009})}\BibitemShut {NoStop}%
\bibitem [{\citenamefont {Wu}(2010)}]{wu-anharmonic-prb}%
  \BibitemOpen
  \bibfield  {author} {\bibinfo {author} {\bibfnamefont {Z.}~\bibnamefont
  {Wu}},\ }\href@noop {} {\bibfield  {journal} {\bibinfo  {journal} {Phys. Rev.
  B}\ }\textbf {\bibinfo {volume} {81}},\ \bibinfo {pages} {172301} (\bibinfo
  {year} {2010})}\BibitemShut {NoStop}%
\bibitem [{\citenamefont {Kraus}\ \emph {et~al.}(2014)\citenamefont {Kraus},
  \citenamefont {Root}, \citenamefont {Lemke}, \citenamefont {Stewart},
  \citenamefont {Jacobsen},\ and\ \citenamefont {Mattsson}}]{Kraus2014}%
  \BibitemOpen
  \bibfield  {author} {\bibinfo {author} {\bibfnamefont {R.~G.}\ \bibnamefont
  {Kraus}}, \bibinfo {author} {\bibfnamefont {S.}~\bibnamefont {Root}},
  \bibinfo {author} {\bibfnamefont {R.~W.}\ \bibnamefont {Lemke}}, \bibinfo
  {author} {\bibfnamefont {S.~T.}\ \bibnamefont {Stewart}}, \bibinfo {author}
  {\bibfnamefont {S.~B.}\ \bibnamefont {Jacobsen}}, \ and\ \bibinfo {author}
  {\bibfnamefont {T.~R.}\ \bibnamefont {Mattsson}},\ }\href@noop {} {\bibfield
  {journal} {\bibinfo  {journal} {\textit{Accepted for Publication, Nature
  Geo.}}\ } (\bibinfo {year} {2014})}\BibitemShut {NoStop}%
\bibitem [{\citenamefont {Croft}(1982)}]{Croft1982}%
  \BibitemOpen
  \bibfield  {author} {\bibinfo {author} {\bibfnamefont {S.~K.}\ \bibnamefont
  {Croft}},\ }\href@noop {} {\bibfield  {journal} {\bibinfo  {journal} {Geol.
  Soc. America, Spec. Pubs.}\ }\textbf {\bibinfo {volume} {190}},\ \bibinfo
  {pages} {143} (\bibinfo {year} {1982})}\BibitemShut {NoStop}%
\bibitem [{\citenamefont {Svendsen}\ and\ \citenamefont
  {Ahrens}(1987)}]{SvendsenMgO}%
  \BibitemOpen
  \bibfield  {author} {\bibinfo {author} {\bibfnamefont {B.}~\bibnamefont
  {Svendsen}}\ and\ \bibinfo {author} {\bibfnamefont {T.~J.}\ \bibnamefont
  {Ahrens}},\ }\href@noop {} {\bibfield  {journal} {\bibinfo  {journal}
  {Geophys. J. R. Astr. Soc.}\ }\textbf {\bibinfo {volume} {91}},\ \bibinfo
  {pages} {667} (\bibinfo {year} {1987})}\BibitemShut {NoStop}%
\bibitem [{\citenamefont {Alfe}(2005)}]{Alfe-MgO-2phase}%
  \BibitemOpen
  \bibfield  {author} {\bibinfo {author} {\bibfnamefont {D.}~\bibnamefont
  {Alfe}},\ }\href@noop {} {\bibfield  {journal} {\bibinfo  {journal} {Phys.
  Rev. Lett.}\ }\textbf {\bibinfo {volume} {94}},\ \bibinfo {pages} {235701}
  (\bibinfo {year} {2005})}\BibitemShut {NoStop}%
\end{thebibliography}
%

\end{document}